\documentclass{iopart}
\usepackage{graphicx}
\usepackage{iopams}
\usepackage{xcolor}
\usepackage{appendix}
\usepackage{subeqnarray}
\usepackage{url}

\newcommand{\iotabar}{\raisebox{-2.5pt}{$\mathchar'26$}\mkern-7.5mu \iota}

\begin{document}
\title[Compressible impurity flow in the TJ-II stellarator]
{Compressible impurity flow in the TJ-II stellarator}
\author{J Ar\'evalo$^1$, J A Alonso$^1$, K J McCarthy$^1$, J L Velasco$^1$, J M Garc\'ia-Rega\~na$^2$ and M Landreman$^3$}
\address{$^1$ Laboratorio Nacional de Fusi{\'o}n, Asociaci\'on EURATOM-CIEMAT, Madrid, Spain}
\address{$^2$ Max-Planck-Institut f\"ur Plasmaphysik, EURATOM-Assoziation, Garching, Germany}
\address{$^3$ Plasma Science and Fusion Center, MIT, Cambridge, Massachusetts, USA}
\ead{juan.arevalo@externos.ciemat.es}
\begin{abstract}
Fully-ionised carbon impurity flow is studied in ion-root, neutral beam heated plasmas by means of Charge Exchange Recombination Spectroscopy (CXRS) in the TJ-II stellarator. Perpendicular flows are found to be in reasonable agreement with neoclassical calculations of the radial electric field. The parallel flow of the impurity is obtained at two locations of the same flux surface after subtraction of the calculated Pfirsch-Schl\"uter parallel velocity. For the medium density  plasmas studied, $\bar{n}_{\rm e}\!\in\!(1.2\!-\!2.4)\times\! 10^{19}$ m$^{-3}$, the measured impurity flow is found to be inconsistent with a total incompressible flow, i.e. $\nabla\cdot{\bf u}_z\ne0$, thus contradicting the usual assumption of a constant density on each flux surface. % Equivalently, the flow is found to be compressible. 
The experimentally observed velocity deviations are compared with the parallel return flow calculated from a modelled impurity density redistribution driven by ion-impurity friction. Although the calculated return flow substantially modifies the incompressible velocity pattern, the modifications do not explain the in-surface variations of impurity parallel mass flow at the precise locations of the CXRS measurements. Small inhomogeneities of the electrostatic potential in a surface are also shown to affect the impurity redistribution but do not provide  a better understanding of the measurements. 
\end{abstract}
\pacs{52.25.Vy, %impurities in plasma 
52.30.-q, % plasma dynamics and flow
52.55.Hc, % stellarators, etc. 
52.70.Kz, % optical measurements
52.25.Dg %plasma kinetic equations
}
\section{Introduction}

% Plasma flows in general
The flow of mass along flux surfaces in magnetically confined plasmas has come to be regarded as an important factor in determining plasma stability, radial transport and performance of these devices. The $E\times B$ flow pattern is of particular importance with regard to transport. Indeed, a sufficiently strong radial velocity shear is generally accepted as reducing turbulence and transport \cite{TerryRevModPhys2000}. The flow patterns of the different species present in the plasma (main ions, electrons and impurity ions) deviate from the $E\times B$ flow and from each other through their different diamagnetic velocity components and parallel force balances. These diamagnetic and parallel flows give rise to currents that are a fundamental part of the stellarator MHD equilibrium in high-beta reactor-relevant plasmas. Parallel currents are generally split into Pfirsch-Schl\"uter (PS) and parallel mass flows. The former arises in response to the compressibility of the perpendicular diamagnetic current and carries no net toroidal current, but can nevertheless cause a radial (Shafranov) shift of magnetic surfaces as the pressure gradient increases. The reduction of this current is a design requirement of modern stellarators because of its detrimental effect on high beta stability and neoclassical transport \cite{WobigPPCF1999}. On the other hand, the bootstrap current carries a net toroidal current\footnote{In the parallel mass flow we therefore include all the relevant parallel forces that determine the parallel flows of the different species, e.g. the NBI-driven currents. It is also noted that, to some extent, the split of the parallel currents is a matter of convention \cite{CoronadoPoFB1992}.} and can potentially change the iota profile, which is of particular importance for island divertor configurations in stellarators and is taken advantage of in tokamak non-inductive scenarios. For these reasons experimental validation of first-principle theory-based models of plasma flows and currents is of considerable importance.
% Our previous results
In reference \cite{ArevaloNF2013} measurements of fully ionised carbon impurity flow were undertaken using Charge Exchange Recombination Spectroscopy \cite{IslerPPCF1994} (CXRS) in low density, Electron Cyclotron Resonance heated plasmas in the TJ-II stellarator. It was verified  that the in-surface variation of the parallel impurity flow  was consistent with an incompressible total flow tangent to flux surfaces. In addition, the  measured perpendicular and parallel mass flows were compared with neoclassical calculations of the radial electric field and ion parallel mass flow, finding a good agreement in those low density plasmas.

% Reported in-out density variations in tokamaks
The CXRS diagnostic in TJ-II makes use of C-VI spectral lines to measure the temperature, density and velocity of C$^{6+}$ ions. Large uncertainties in carbon density profile measurements arise  in TJ-II from the lack of calibration of the optical paths inside the vacuum chamber. Therefore, to study whether the density is constant on flux surfaces the approach taken is to examine the spatial variation of the flows, as in references \cite{MarrPPCF2010,PuterichNF2012}. On the other hand, profiles of C$^{6+}$ temperature and parallel flow are routinely measured in TJ-II \cite{ArevaloNF2013,CarmonaPFR2008}. Now, because of their high collision frequency with main ions, the temperature and parallel mass flow of impurities  are generally taken to be a proxy for the main ion temperature and parallel flow. However, as noted in e.g. \cite{Helander_PoP1998}, the strong collisional character of high-Z impurities can cause impurity density variations within a surface to be comparable with their mean value on the surface. This ratio scales as $\tilde{n}_z/n_z \sim \delta_i \hat{\nu}_{ii}Z^2$, where $\delta_i=\rho_i/L_\perp$ and $\hat{\nu}_{ii} = L_\|/\lambda_{i}$ are the normalised ion gyro-radius and collisionality, respectively ($\rho_i$: ion Larmor radius, $\lambda_i$: ion mean-free path, $L_\perp$: radial scale length associated with pressure profiles, $L_\parallel$: connection length). The relative variations of the impurity density within a surface can be of order unity, $\tilde{n}_z/n_z\sim\mathcal{O}(1)$, even with $\delta_i\ll 1$ still a valid expansion parameter. Indeed, poloidal  density variations of high-Z impurities have been observed in several tokamak devices (see reference \cite{ReinkePPCF2012} and references therein). These  in-surface density variations cause the  impurity parallel mass flow to  deviate from the ion parallel mass flow as the parallel pressure gradient drive a return flow. Furthermore, the return flows need not preserve an incompressible pattern as particle conservation $\nabla\cdot (n_z\mathbf{u}_z) = 0$ does not reduce to flow incompressibility $\nabla\cdot\mathbf{u}_z = 0$ for a non-constant impurity density $n_z$. Recent experiments using the CXRS technique on light impurities (B$^{5+}$ and C$^{6+}$) have reported a poloidal asymmetry of the impurity parallel flow at the pedestal region of the Alcator C-Mod \cite{MarrPPCF2010} and AUG \cite{PuterichNF2012} tokamaks, thus indicating a substantial poloidal density variation of light-Z impurities from particle conservation. Such a poloidal redistribution of the impurities was confirmed later with direct impurity density measurements \cite{ChurchillAPS2012,ViezzerEPS2013}. In stellarators, a poloidal variation of the C$^{6+}$ density and poloidal velocity was observed in the core region of outward shifted plasmas in CHS \cite{NishimuraPoP2000}.

% impurity radial transport
While such behaviour of the impurity flows limits the capability of CXRS on fully ionized impurity ions to directly track the bulk flow (arguably its intended purpose), the study of impurity density asymmetries and flow variations provides valuable information on impurity-ion friction, which is expected to be an important mechanism of impurity radial transport (see, e.g. reference \cite{Braun_Helander_ConfSeries2010}). Impurity dynamics, and particularly their radial transport and accumulation in  the core of fusion plasmas, is of utmost importance for fusion viability both in tokamaks and stellarators, as impurities can produce unacceptable energy losses through radiation as well as diluting the fusion reactants. Nevertheless, the radial transport of impurities displays considerable complexity which often has eluded a theoretical explanation. Classical examples are the impurity screening shown by the High-Density H-mode observed in W7-AS \cite{McCormickPRL2002} and the impurity hole in LHD plasmas \cite{YoshinumaNF2009}. The experimental validation of theoretical models of impurity density redistribution within a flux surface is of considerable importance as it provides some indirect validation of the model predictions on impurity radial transport \cite{Regana2012}.

% What do we do here 
In this work, CXRS measurements of C$^{6+}$ flows in Neutral Beam Injection (NBI) heated, ion-root plasmas of the TJ-II stellarator are presented. As indicated in reference \cite{ArevaloNF2013} significant and reproducible deviations in the measured impurity flow from an incompressible pattern are observed as density increases, which points to a redistribution of impurity density within the flux surfaces as those observed in \cite{MarrPPCF2010,PuterichNF2012} although it appears in more internal regions of the plasma, as in \cite{NishimuraPoP2000}. We present these flow measurements and compare the observed deviations with the parallel return flow from a modelled impurity density redistribution driven by ion-impurity friction \cite{Helander_PoP1998}.  Such a friction model was adapted to a general stellarator geometry with the bulk ions in the Pfirsch-Schl\"uter regime of collisionality in reference \cite{Braun_Helander_ConfSeries2010} and is extended here for main ions in the plateau regime,  provided the ion temperature gradient is small (a plausible assumption for the plasmas under consideration). The calculated return flow substantially modifies the incompressible velocity pattern, being comparable to the impurity parallel PS flow. However, it is shown that the calculated modifications at the precise locations of the CXRS measurements are small in comparison to the measured in-surface variations of impurity parallel flow and in the opposite direction for most cases. The inclusion of inertial and parallel electric field forces in the parallel momentum balance does not provide a better understanding of the experimental observations.

% Outline
This paper is organised as follows: in section \ref{sec:experimental_setup}, the diagnostic set-up and geometry are presented together with the methodology used to relate the flow fields to the CXRS velocity measurements through the appropriate geometric quantities. In section \ref{sec:results} the impurity flow measurements and their compressible asymmetries are described. These asymmetries are compared to the outputs of an ion-impurity friction model in section \ref{sec:discussion}, where modifications to the impurity flow incompressible pattern,  caused by an in-surface impurity density variation, are  detailed. In section \ref{sec:extensions} the validity of the friction model is examined and the impurity parallel force balance is extended to account for inhomogeneities of the electrostatic potential within a magnetic surface.  Finally, conclusions are drawn in section \ref{sec:conclusions}.

%%%%%%%%%%%%%%%%%%%%%%%%%%%%%%%%%%%%%%%%%%%%%%%%%%%%%%%%%%%%%%%%%%%%%%%%%%%%%%
\section{Diagnostic set-up and data analysis}\label{sec:experimental_setup}
%%%%%%%%%%%%%%%%%%%%%%%%%%%%%%%%%%%%%%%%%%%%%%%%%%%%%%%%%%%%%%%%%%%%%%%%%%%%%%
The TJ-II CXRS diagnostic set-up and data analysis have been reported previously in reference~\cite{ArevaloNF2013}. Here, the CXRS process of interest involves electron capture from accelerated hydrogen by fully ionized carbon ions into a highly excited state  of C$^{5+}$, followed by spontaneous decay via photon emission, i.e. the C VI line at 529.07 nm ($n=8\rightarrow 7$). For this, a compact Diagnostic Neutral Beam Injector (DNBI) provides a 5 ms long pulse of neutral hydrogen accelerated to $30$ keV. Its $1/e$-radius at focus is $21$ mm \cite{CarmonaRSI2006}.

\begin{figure}[htb]
\centering
\includegraphics[width=7 cm]{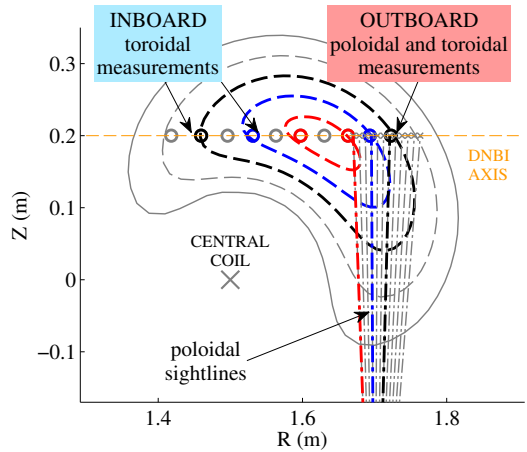}
\caption{\label{fig:sight-lines}Schematic diagram of CXRS diagnostic sightlines with a poloidal cut of several magnetic surfaces of TJ-II. %Toroidal views go into the page. 
The inboard and outboard regions of measurement are highlighted. The magnetic surfaces in which poloidal and toroidal outboard measurements are coincident, $\rho\sim 0.2, 0.4$ and $0.6$, are coloured in red, blue and black, respectively.}
\end{figure} 

A schematic layout of the diagnostic sightlines is presented in figure~\ref{fig:sight-lines}, together with a poloidal cut of several magnetic surfaces of TJ-II. The plasma minor radius region spanned by nearly symmetric poloidal views is $\rho\in(0.25,0.85)$ in the magnetic configurations studied in this work. Here the normalised radius is defined as $\rho\equiv\sqrt{V/V_0}$, where $V$ and $V_0$ are the volumes enclosed by the surface of interest and the last closed magnetic surface, respectively. In the figure only the bottom poloidal array is presented for clarity (see reference~\cite{ArevaloNF2013} for details). On the other hand, the toroidal fibres cover both sides of the magnetic axis, from $\rho \sim-0.75$ to $\rho=0.6$ at 10 locations (in figure~\ref{fig:sight-lines} the toroidal sightlines, plotted as grey circles, go outside the page). The region in which both poloidal and toroidal measurements are taken is labelled as outboard, while the zone where only toroidal measurements are made is labelled as inboard. The nomenclature $\rho\ge0$ (outboard) and $\rho\le0$ (inboard) is also utilized to define these regions. 

In the outboard region, poloidal and toroidal fibres view the same surfaces at $\rho\sim 0.2, 0.4$ and $0.6$, see figure~\ref{fig:sight-lines}. Therefore, the 2D-flow velocity is completely determined at these locations. The redundant inboard-toroidal measurements have been recently used to demonstrate that, in low density TJ-II plasmas (with line-averaged electron densities $\bar{n}_e\le 10^{-19}$ m$^{-3}$), impurity rotation is incompressible and follows neoclassical theory~\cite{ArevaloNF2013}. The methodology used to verify incompressibility is reviewed next, and extended to account for a compressible velocity pattern. 

%------------------------------------------------
\subsection{Spatial variation of the flow}\label{sec:flow_structure}
%------------------------------------------------
From the radial force balance for a single species $s$ it follows that the perpendicular velocity is given by the ${\bf E}\times{\bf B}$ and diamagnetic flows~\cite{HintonRMP1976},
\begin{equation*}
{\bf u}_{s\perp}=\frac{{\bf B}}{B^2}\times
\left(\nabla\Phi+\frac{1}{n_s q_s}\nabla p_s\right)+\mathcal{O}(\delta_s^2v_s).
\end{equation*}
where $n_s$, $q_s\!=\!e Z_s$ and $v_s\!=\!\sqrt{2T_s/m_s}$ are the density, charge and thermal velocity of the species $s$.
The electric potential $\Phi$ and pressure $p_s$ are approximately constant on flux surfaces, so
\begin{equation}\label{eq:u_perp}
{\bf u}_{s\perp}=E_s\frac{{\bf B}\times\nabla\rho}{B^2},\quad
E_s=\frac{{\rm d}\langle\Phi\rangle}{{\rm d}\rho}+\frac{1}{\langle n_s\rangle q_s}\frac{{\rm d} \langle p_s\rangle}{\rm{d}\rho}.
\end{equation}
Here, poloidal variations of the particle density are neglected against the radial pressure gradient. Furthermore, when impurities are considered the diamagnetic  term is indeed taken as a small correction to the dominant $E\times B$ flow, due to the $Z_s^{-1}$ factor.

The form of the parallel flow is obtained from the (steady-state) $s$ number conservation, $\nabla\cdot(n_s{\bf u}_s)=0$. If density is constant on flux surfaces then $\nabla\cdot{\bf u_s}=0$ and a local parallel flow (Pfirsch-Schl\"uter) must compensate for the compression of the perpendicular flows. The general expression of the parallel flow is then (see e.g. \cite{ArevaloNF2013,SugamaPoP2002})
\begin{equation}
\label{eq:general_parallel_flows}
{\bf u}_{s\parallel}=\left(E_s h + \Lambda_s\right){\bf B},
\end{equation}
with the function $h(\rho,\theta,\phi)$ ($\theta,\phi$: poloidal and toroidal angles) satisfying 
\[
{\bf B}\cdot\nabla h = \frac{2}{B^2}{\bf B}\times\nabla \rho\cdot\nabla\left(\ln{B}\right).
\]
The integration constant for $h$ is fixed by the condition $\langle hB^2\rangle=0$. With this choice the flux-constant $\Lambda_s(\rho)$ in \eref{eq:general_parallel_flows} is given by $\Lambda_s\equiv \langle{\bf u}_s\cdot{\bf B}\rangle/\langle B^2\rangle$,
where $\langle \cdot\rangle$ denotes a flux-surface average. The first term on the right of equation \eref{eq:general_parallel_flows}, ${\bf u}_{\rm PS}=E_sh{\bf B}$, is the well-known Pfirsch-Schl\"uter flow. Hereafter, $\Lambda_s {\bf B}$ is referred as the parallel mass flow of the $s$ species, i.e. the parallel flow without the PS contribution.

As mentioned in the introduction, the collisional character of high-$Z$ impurity ions can give rise to large variations of the impurity density within a flux surface, making $n_z/\langle n_z\rangle \sim 1$. In this case the function $\Lambda_z$ above is no longer a flux function. Then it is convenient to define an impurity parallel return flow as $\Lambda(\rho,\theta,\phi) = \Lambda_z(\rho,\theta,\phi) - \Lambda_i(\rho)$, which is associated with parallel gradients of the impurity density. % and carries no net parallel flow $\langle \Lambda B^2\rangle = 0$.
With this particular choice impurity flows are written as the sum of an incompressible flow, 
\begin{equation}\label{eq:incompressible_flow}
{\bf u}_{z0} =  E_z\frac{{\bf B}\times\nabla\rho}{B^2} + \left(\Lambda_i+E_z h\right){\bf B},
\end{equation}
plus the return flow, $\Lambda {\bf B}$, which compensates for the impurity density redistribution, i.e.
\begin{equation}\label{eq:compressible_flow}
{\bf u}_z = {\bf u}_{z0} + \Lambda(\rho,\theta,\phi){\bf B}.
\end{equation}
Note: this velocity field is the same as that given by equations \eref{eq:u_perp} and \eref{eq:general_parallel_flows}, but with $\Lambda_z(\rho,\theta,\phi)=\Lambda(\rho,\theta,\phi)+\Lambda_i(\rho)$.
%------------------------------------------------
\subsection{Data analysis}
%------------------------------------------------
For data analysis purposes it is convenient to define dimensionless vectors, ${\bf f}$ and ${\bf g}$,  and flux-surface constants with units of velocity. To this end, the incompressible impurity flow pattern in \eref{eq:u_perp} and \eref{eq:general_parallel_flows} can be recast  as ${\bf u}_z={\bf f}U_\perp+{\bf g}U_{\rm b}$, where the flux-surface averaged (FSA) flows are~\cite{ArevaloNF2013}
\begin{eqnarray*}
U_\perp(\rho)&\equiv&\frac{E_{\rm r}}{\langle B\rangle}-
\frac{1}{eZ\langle n_z\rangle}\frac{{\rm d}\langle p_z\rangle}{{\rm d}\rho}
\frac{\langle |\nabla \rho|\rangle}{\langle B\rangle},\label{eq:Uperp}\\
U_{\rm b}(\rho)&\equiv&\frac{\langle{\bf u}_z\cdot{\bf B}\rangle}{\langle B\rangle}=\Lambda_z(\rho)\frac{\langle B^2\rangle}{\langle B\rangle},
\end{eqnarray*}
and $E_{\rm r}(\rho)\equiv -\langle |\nabla \rho|\rangle {\rm d\langle\Phi\rangle}/{\rm d\rho}$ is the FSA radial electric field. The variation of the velocity field within the surface has been stored in the dimensionless vectors ${\bf f}$ and ${\bf g}$,
\begin{eqnarray*}
\label{eq:dimensionless_factors}
{\bf f}&=&-\frac{\langle B\rangle}{\langle|\nabla \rho|\rangle}
\left(\frac{{\bf B}\times\nabla\rho}{B^2}+h{\bf B}\right),\\
{\bf g}&=&\frac{\langle B\rangle}{\langle B^2\rangle}{\bf B}.
\end{eqnarray*}
Note that the vector ${\bf f}$ has both perpendicular and parallel components, the latter coming from the PS contribution. A first set of FSA flows, $(U_\perp,U_{\rm b})^{\rm out}$, is obtained from the poloidal and toroidal velocities measured in the outboard region, see figure \ref{fig:sight-lines}. Then, the toroidal inboard measurements are combined with the poloidal outboard ones to calculate a second set $(U_\perp,U_{\rm b})^{\rm in}$, which is compared with the outboard FSA flows. In the case of an incompressible flow, since $U_\perp$ and $U_{\rm b}$ are flux-constants, the outboard and inboard measurements should give the same results.

On the contrary, for a compressible flow like that of equation~\eref{eq:compressible_flow}, the two measurements of $(U_\perp,U_{\rm b})$ will not match. In order to parametrize this effect the differences in the measured parallel flow $\Lambda_z(\rho,\theta,\phi){\bf B}$ are studied. To this end, the velocity field \eref{eq:compressible_flow} is expressed as
\begin{equation}\label{eq:CXRS_velocity_field_compressible}
{\bf u}_z={\bf f}U_\perp+\Lambda_z{\bf B}.
\end{equation}
Then, the quantity $\Lambda_z(\rho,\theta,\phi)$ is obtained after projecting this flow onto toroidal sightlines,
\begin{equation}\label{eq:lambda_z}
\Lambda_z = \frac{{u}_{\rm t}-{f}_{\rm t} U_\perp}{{ B}_{\rm t}}.
\end{equation}
Here, the subindex ${\rm t}$ indicates the projection of a vector (${\bf u},~{\bf f}$ and ${\bf B}$) in the toroidal viewing direction,  ${\bf e}_{\rm t}$. The measured toroidal flow is therefore $u_{\rm t}={\bf u}\cdot{\bf e}_{\rm t}$. Note that the perpendicular flow in \eref{eq:CXRS_velocity_field_compressible} is assumed to be a flux-constant. Thus, it can be evaluated in the outboard region, $U_\perp(\rho) \equiv U_\perp^{\rm Out}(\rho)$, where both poloidal and toroidal velocities are measured, see figure \ref{fig:sight-lines}.
Finally, the differences in the parallel mass flow (divided by the local magnetic field strength) are
\begin{equation}\label{eq:differences_lambda_z}
\Delta\Lambda_z = \Delta\left(\frac{u_{\rm t}}{B_{\rm t}}\right)-
U_\perp\Delta\left(\frac{f_{\rm t}}{B_{\rm t}}\right),
\end{equation}
where $\Delta (X)\equiv X^\mathrm{In}-X^\mathrm{Out}$. Note that the differences in the impurity parallel mass flow equal those of the impurity return flow defined in equation \eref{eq:compressible_flow}, i.e. $\Delta\Lambda_z\equiv\Delta\Lambda$. Therefore, if flows were incompressible, $\Delta\Lambda_z=0$.

Since the possibility of an impurity compressible flow is studied, carbon density measurements are of great interest. These are estimated in TJ-II after assuming that the efficiency of the optical system is proportional to the inverse of integrated calibration signal (see equation (2) in~\cite{ArevaloNF2013}). However, as the lamp used for  spectrograph calibration is located  between the light collection lens and fibre bundle (which are external to the vacuum chamber), the influence of the internal mirror for the toroidal sightlines, the vacuum  viewports or the focusing lenses are not included in the calibration, thereby introducing significant uncertainties into the carbon density profile measurements. Therefore, these measurements are not used in this paper.
\subsection{Optical alignment}
The procedure for aligning the diagnostic is detailed in section 2.1 of reference~\cite{ArevaloNF2013}. Since $\beta$ in the present experiments is low ($\beta\leq0.5~\%$), the location of the flux surfaces is known accurately from the vacuum field. In this regard, mapping measurements along flux surfaces between the outboard and inboard sides is less error-prone in a stellarator than in a tokamak, where reconstruction of the magnetic equilibrium can introduce significant uncertainty. Consequently, the inboard and outboard measurements are directly mapped to flux coordinates  using the known magnetic field geometry, and in contrast to the tokamak experiments \cite{MarrPPCF2010,PuterichNF2012, ViezzerEPS2013}, no additional relative shift between the inboard and outboard measurements is needed to align the carbon temperature profiles.

Moreover, an additional check to confirm the goodness of the toroidal $\rho$ mapping has been made for this work. For this, the CXRS grating (set at 529 nm for C$^{6+}$ measurements \cite{CarmonaRSI2006}) in the spectrograph was exchanged for one centred at 656.2 nm to measure H$_\alpha$ emission from the beam. Thus, by injecting the DNBI beam into the vacuum chamber without magnetic fields, spectra with Doppler-shifted H$_\alpha$ line emission from the beam were collected and analysed. Then, by determining the Doppler shift of the H$_\alpha$ beam light for each sight line, the corresponding beam velocity is calculated without corrections for the beam to sight line angle. Knowing the beam energy, the beam to sight lines angles can be determined and the beam/sightlines intersection points can be determined. These intersection points  are  compared with the values obtained using the illumination method described in~\cite{ArevaloNF2013} showing that the uncertainties in the alignment of toroidal fibres are $\sim\pm 3$ mm. Note: the separation between toroidal sighlines, $\geq 3$ cm, is fixed by the fibre bundle and focusing lens.

%%%%%%%%%%%%%%%%%%%%%%%%%%%%%%%%%%%%%%%%%%%%%%%%%%%%%%%%%%%%%%%%%%%%%%%%%%%%%%%%%%%%%%%%%%%%%%%%%%
\section{Experimental results}\label{sec:results}
%%%%%%%%%%%%%%%%%%%%%%%%%%%%%%%%%%%%%%%%%%%%%%%%%%%%%%%%%%%%%%%%%%%%%%%%%%%%%%%%%%%%%%%%%%%%%%%%%%
In this work, two close magnetic configurations are considered: $100\_44\_64$ and $100\_40\_63$. Here, the nomenclature reflects currents in the central, helical and vertical coils, respectively. The vacuum rotational transform, $\iotabar$, covers the range $1.55\le\iotabar\le 1.65$ and $1.509\le\iotabar\le 1.608$, and the volumes are $1.098$ and $1.043$ m$^3$, respectively. These two configurations have been studied in references \cite{VelascoPPCF2011} and \cite{VelascoPPCF2012} from the neoclassical point of view. For similar plasma profiles and momentum input, no qualitative differences are predicted in the flows within the surface. The plasmas presented here are heated by one of the two tangential NBI, either in the direction of the magnetic field (co-injection), or in the opposite direction (counter-injection). The line averaged  densities scanned in this paper cover the range $\bar{n}_e\in (1.2-2.4)\times 10^{19}$ m$^{-3}$. 

\begin{figure}[!h]
\centering
\includegraphics[width=7 cm]{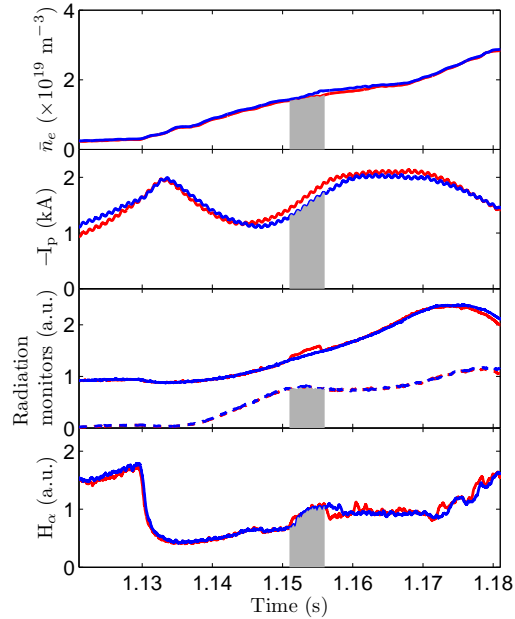}
\caption{\label{fig:time_evolution}Time evolution of two similar plasma discharges. In red, shot$\#32577$ with a DNBI pulse and in blue, shot$\#32576$ without DNBI, used to remove the C$^{5+}$ passive contribution. From top to bottom, time traces of: line averaged electron density, $\bar{n}_e$; plasma current (reversed), $-{\rm I_p}$; radiation monitors: bolometer (solid line) and C$^{4+}$ (dashed); $H_\alpha$ signal. The DNBI injection is shown in grey.}
\end{figure} 
Time traces of a representative plasma discharge, $\#32577$, in the $100\_44\_64$ configuration are shown in figure~\ref{fig:time_evolution}. The evolution of the reference discharge ($\#32576$) used to remove background C$^{5+}$ emission is in blue, whilst the one in which the DNBI was fired, $\#32577$, is in red. The time interval in which the DNBI is injected is plotted as a grey shadow. The NBI heating causes an increase in the line-averaged density and radiation. The radiation monitors in figure \ref{fig:time_evolution} correspond to a bolometer signal (whose view-line intersects the DNBI path) and a C$^{4+}$ monitor, shown as solid and dashed lines, respectively. A small increase is observed in the bolometer signal for discharge $\#32577$, which corresponds to photon excitation induced by the DNBI. Finally, the plasma current, ${\rm I_p}$, is negative corresponding to counter-injection (a co-injection reverses the sign of the current). The small oscillation observed in the plasma current is produced by small variations in the current of the external coils. The good reproducibility of the two discharges shown in figure \ref{fig:time_evolution} is representative of the data set used in this work and allows an accurate subtraction of the background C$^{5+}$ emission from the active DNBI discharge.

\begin{figure}[!h]
\centering
\includegraphics[width=7 cm]{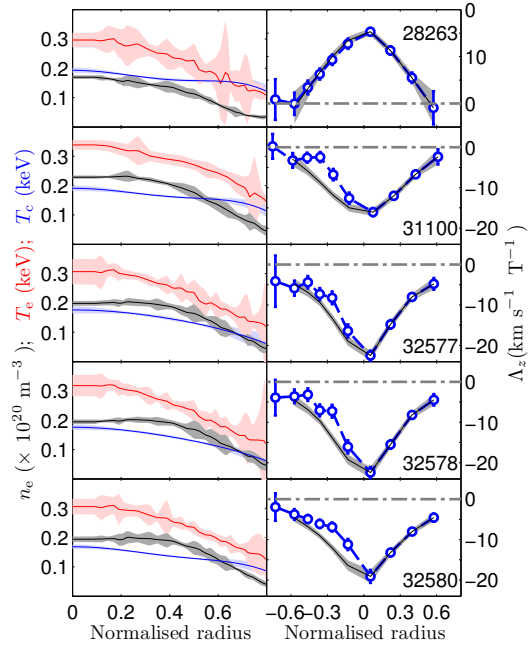}
\caption{\label{fig:plasma_profiles} Left: profiles of the electron density ($n_{\rm e}$, in gray) and temperature ($T_{\rm e}$, in red), together with carbon temperature profiles ($T_{\rm c}$, in blue). Right: measured profile of $\Lambda_z$, in blue, and the incompressible expectation  extrapolated from the outboard measurements, in grey. The discharge $\#28263$, in which flows were demonstrated to be incompressible~\cite{ArevaloNF2013}, is included here as a reference. The discharges $\#31100$, $\#32577$, $\#32578$ and $\#32580$, heated with one NBI in counter-injection (consistent with $\Lambda_z<0$) are presented to highlight the reproducibility of the departure from incompressibility observed in $\Lambda_z$. The discharge $\#31100$ was performed in the configuration $100\_40\_63$, and six months before the other three, measured in the $100\_44\_64$ magnetic configuration.}
\end{figure}
Electron density and temperature profiles ($n_{\rm e}$ and $T_{\rm e}$, respectively) are measured by the Thomson scattering diagnostic~\cite{milligen2011bayes}, see figure ~\ref{fig:plasma_profiles}. The discharges in the figure were performed in the $100\_44\_64$ magnetic configuration, except discharge $\#31100$ (configuration $100\_40\_63$). In all the discharges considered in this paper, the electron temperature profile is approximately parabolic, with $T_{\rm e}(0)\approx 300$ eV, whilst the carbon temperature profile is rather flat, with $100\leq T_{\rm c}\leq 200$ eV (it is assumed that main ion  and impurities are in thermal equilibrium). In the right column of figure~\ref{fig:plasma_profiles} the measured C$^{6+}$ parallel mass flow  profile $\Lambda_z$ is shown in blue (see section \ref{sec:experimental_setup} for an explanation of the extraction of this flow component from the experimental measurements). The direction of the bulk toroidal flow is mainly determined by the NBI momentum injection. The incompressible expectation  extrapolated from the outboard measurements, i.e. $\Lambda_z=\Lambda_z(\rho)= \Lambda_z^{\rm Out}$, is shown in grey. The discharge $\#28263$ is heated by the co-NBI injector and shows a lower line-averaged density ($\bar{n}_{\rm e}=1.2\times 10^{19}$ m$^{-3}$). For this low-density NBI discharge flows were shown to be incompressible in reference~\cite{ArevaloNF2013}, and is included here  as a reference. The measured $\Lambda_z$ profile departs from the incompressible expectation in discharges $\#31100$, $\#32577$, $\#32578$ and $\#32580$. The reproducibility of the $\Lambda_z$ profile for similar discharges in terms of $n_{\rm e}, T_{\rm e}$ and $T_{\rm c}$ profiles, but otherwise distant in time and impurity content, reinforces the reproducibility of the observed flow deviations. The general tendency observed in the experimental database, with few exceptions, is that the inboard parallel flow is more positive than the outboard one, and thus, the in-out differences in the parallel mass flow, $\Delta\Lambda_z$ from equation \eref{eq:differences_lambda_z}, are always positive. This observation is nearly independent on the magnetic configuration and the direction of injection of the heating NBIs.% $\Delta\Lambda_z\!>\!0$. 

\begin{figure}[!h]
\centering
\includegraphics[width=6 cm]{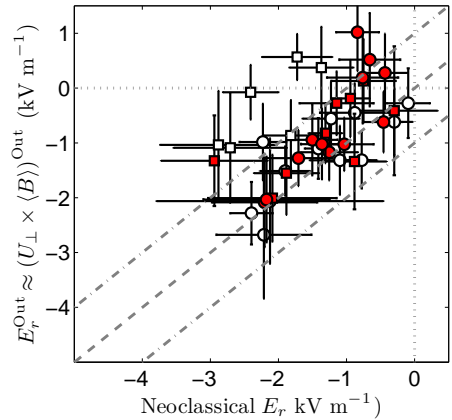}
\caption{\label{fig:Er_nbi}Comparison of the experimentally measured radial electric fields for the outboard region, $E_r^{\rm Out}$, with the corresponding neoclassical values for several TJ-II discharges. Here, circles and squares represent data from the $100\_44\_64$ and $100\_40\_63$ magnetic configurations, respectively. Red points indicate NBI in counter-${\bf B}$ direction (consistent with $\Lambda_z<0$) while white points indicate co-injection. Note that the impurity diamagnetic term is not included in computing the measured $E_r$ (see text).}
\end{figure}

As indicated in section \ref{sec:experimental_setup}, toroidal and poloidal view lines of the CXRS system overlap at three locations ($\rho \approx 0.2, 0.4$ and $0.6$) on the outboard side of the DNBI path. This enables unambiguous determination of the perpendicular and parallel flow components at those locations, relying only on the assumption of a small radial flow component compared to perpendicular and parallel flows. The perpendicular impurity flow component is expected to be dominated by the $E\times B$ flow because of the $1/Z$ factor of the diamagnetic flow. Figure \ref{fig:Er_nbi} shows the comparison of this experimental approximation to the radial electric field with the neoclassical expectations, calculated as in~\cite{VelascoPPCF2012}. The database shown here is comprised of $12$ discharges. Data taken in the $100\_44\_64$ and $100\_40\_63$ magnetic configurations are presented as circles and squares, respectively. Note that the impurity diamagnetic term is not included in the calculation of the experimental radial electric field, because of the uncertainties in determining the carbon density profile in TJ-II. Nevertheless, a rough estimate of the diamagnetic contribution (obtained from the CXRS signals while ignoring the calibration deficiencies mentioned in section \ref{sec:experimental_setup}) typically results in absolute values $\leq 1$ kV m$^{-1}$ at $\rho=0.6$, with little or no impact for more internal regions, as expected from the $1/Z$ dependence. This estimation  is consistent with the main-ion diamagnetic velocities measured,  $|v_{{\rm diam},i}|\leq 4$ km s$^{-1}$ for $|\rho|\le 0.6$.

Despite this uncertainty in the estimated radial electric field,  we note that the radial electric field does not enter any of the expressions for the impurity flows alone (sections~\ref{sec:flow_structure} and \ref{sec:discussion}), but rather in combination with the diamagnetic component as the total perpendicular flow. Such a velocity component is provided by the overlapping CXRS velocity measurements through geometric factors only and is not subjected to such uncertainties.

%%%%%%%%%%%%%%%%%%%%%%%%%%%%%%%%%%%%%%%%%%%%%%%%%%%%%%%%%%%%%%%%%%%%%%%%%%%%%%%%%%%%%%%%%%%%%%%%%%
\section{Friction-driven impurity density redistribution}\label{sec:discussion}
%%%%%%%%%%%%%%%%%%%%%%%%%%%%%%%%%%%%%%%%%%%%%%%%%%%%%%%%%%%%%%%%%%%%%%%%%%%%%%%%%%%%%%%%%%%%%%%%%%
In order to study the measured  parallel mass flow deviations from an incompressible pattern, the continuity equation
\begin{equation}
\label{eq:continuity_equation}
{\bf B}\cdot\nabla\left(\frac{n_z u_{z\parallel}}{B}\right)=-E_z{\bf B}\times\nabla \rho\cdot\nabla\left(\frac{n_z}{B^2}\right),
\end{equation}
and the impurity parallel force balance
\begin{equation}
\label{eq:impurity_mom_balance}
T_z\nabla_\parallel n_z = R_{z\parallel},
\end{equation}
need to be solved for the unknown functions $n_z$ and $\Lambda_z$. Here, $R_{z\parallel}$ is the parallel friction on the impurities. The inclusion of other forces in \eref{eq:impurity_mom_balance} (namely the impurity inertia and the parallel electric field) is described and evaluated in section \ref{sec:extensions}, while the impurity parallel viscosity is neglected against the parallel impurity pressure gradient, $\nabla_\parallel p_z$ \cite{Helander_PoP1998}. As also shown in reference \cite{Helander_PoP1998}, the strong ion-impurity energy equilibration keeps the impurity temperature close to the ion one and thus $T_z=T_i(\rho)$.

In the axisymmetric tokamak case, the impurity continuity equation \eref{eq:continuity_equation} yields an algebraic relationship between the parallel impurity flow and the impurity density (see e.g. \cite{Landreman_PoP2011}), unlike in a stellarator geometry~\cite{Braun_Helander_ConfSeries2010}, where the coupled partial differential equations
\begin{subeqnarray}
\label{eq:continuity_momentum}
\slabel{eq:continuity_equation_simplified}
{\bf B}\cdot\nabla(n\Lambda)&=&-{\bf u}_{z0}\cdot\nabla n,\\
\slabel{eq:parallel_momentum_simplified}
{\bf B}\cdot\nabla \ln{n}&=&\gamma_{\rm f} B^2\left(A_{\rm i}h+B_{\rm i}- \Lambda\right).
\end{subeqnarray}
need to be solved consistently for $\Lambda$ and $n\equiv n_z/\langle n_z\rangle$, with the solubility condition $\langle \Lambda B^2\rangle=B_i(\rho)\langle B^2\rangle$, see the \ref{sec:appendix}. The compressible pattern \eref{eq:compressible_flow} has been used to express the continuity equation \eref{eq:continuity_equation} in its form \eref{eq:continuity_equation_simplified}. In addition, a flux-constant friction coefficient $\gamma_{\rm f}(\rho)$ and thermodynamic forces $A_{\rm i}(\rho)$ and $B_{\rm i}(\rho)$ have been defined in \eref{eq:parallel_momentum_simplified} as
\begin{subeqnarray}
\label{eq:flux_const_friction}
\slabel{eq:friction_centrif_coeff}
\gamma_{\rm f}&\equiv&\frac{m_iZ^2}{T_{\rm i}\tau_{ii}},\\
\slabel{eq:thermodynamic_force_A}
A_{\rm i}&\equiv&\frac{T_{\rm i}}{e}\frac{{\rm d}\ln n_{\rm i}}{{\rm d}\rho}-\frac{1}{2e}\frac{{\rm d}T_{\rm i}}{{\rm d}\rho},\\
\slabel{eq:thermodynamic_force_B}
B_{\rm i}&\equiv&-\frac{3}{5}\frac{\langle {\bf q}_{\rm i}\cdot {\bf B}\rangle}{p_{\rm i}\langle B^2\rangle},
\end{subeqnarray}
with $\tau_{ii}=3(2\pi)^{3/2}\varepsilon_0^2m_i^{1/2}T_i^{3/2}/(n_{\rm i}e^4\ln{\Lambda})$ the ion self-collision time and ${\bf q}_i$ the ion heat flow. In deriving expression \eref{eq:parallel_momentum_simplified} trace impurities are considered, $\sum n_zZ^2\ll n_i$, and so the parallel friction on the impurities is approximated by that exerted by main ions, i.e. $R_{z\|}\approx R_{zi\|}=-R_{iz\|}$. The ion-impurity collision operator is modelled with a Lorentz pitch-angle scattering operator plus a term guaranteeing momentum conservation \cite{Helander_PoP1998}. Finally, no assumption is made on bulk ion's collisionality since its  distribution function is expanded by Legendre and Laguerre polynomials, as is customary in the so-called moment approach to neoclassical transport \cite{SugamaPoP2002,HelanderCollisions}. Here, the so-called $13$ M approximation \cite{SugamaPoP2010} is adopted, i.e. contributions from $j\!>\!1$ Laguerre components are neglected,  see the \ref{sec:appendix}.

The ion-impurity parallel friction is studied first in the next subsection. The effect of a parallel electric field and impurity inertial forces are considered in  section~\ref{sec:extensions}. We anticipate here that the parallel momentum balance %in \eref{eq:parallel_momentum_simplified} 
is dominated by the friction force and that the general behaviour of the solutions is to display $\Delta\Lambda_z < 0$, in contrast with the measured in-out variation. This can be heuristically understood by noting that the differences in the Pfirsch-Schl\"uter flow, $A_ihB$ in equation~\eref{eq:parallel_momentum_simplified}, drive the impurity density redistribution. As a consequence of the in-surface density variation, an impurity return flow $\Lambda B$ is established (equation \eref{eq:continuity_equation_simplified}) which must act to reduce the overall ion-impurity friction so that the density redistribution is not further amplified. Since the term $A_ih$ on the RHS of equation~\eref{eq:parallel_momentum_simplified} is more negative at the inboard side, the return flow $\Lambda$ tends to behave similarly. %Preliminary calculations including the inhomogeneity of the potential are discussed also in  section~\ref{sec:simulation_results}

%--------------------------------------------
\subsection{Calculation of the friction-driven impurity redistribution}\label{sec:simulation_results}
%--------------------------------------------
The two coupled equations~\eref{eq:continuity_equation_simplified} and \eref{eq:parallel_momentum_simplified}  can be recast as a second order partial differential equation for the unknown function $n(\rho, \theta, \phi)$ (see \ref{sec:appendix}). The radial coordinate is a parameter in those equations which involve angular derivatives only. The required inputs are the main ion  parameters (temperature $T_{\rm i}$, density $n_{\rm i}$, parallel mass flow $\Lambda_{\rm i}$ and flux-surface averaged parallel heat flow $\langle {\bf q}_{\rm i}\cdot {\bf B}\rangle$) together with the impurity perpendicular flow. The CXRS and Thomson scattering systems provide measurements of these parameters, except for the ion parallel mass and heat flows. The latter is calculated with DKES~\cite{hirshman1986dkes}, complemented with momentum correction techniques~\cite{massbergPoP2009}. On the other hand, the measured $\Lambda_z$ in the outboard region is used as a first guess for the ion parallel mass flow to solve the differential equations, $\Lambda_i^{(0)} = \Lambda_z^{\rm Out}$, since the external input of momentum is not included in the DKES $\Lambda_i$ calculations \cite{VelascoPPCF2011}. A new guess for the main ion parallel flow is then obtained upon subtraction of the calculated  impurity-ion flow difference, $\Lambda^{(0)}$ in our notation, i.e. $\Lambda_i^{(1)} = \Lambda_z^{\rm Out} - \Lambda^{(0)}$. Note that at every step momentum conservation is imposed, i.e. $\langle \Lambda B^2\rangle=B_i(\rho)\langle B^2\rangle$ from equation \eref{eq:parallel_momentum_simplified}. The iteration of this process leads to a solution for the impurity flow that matches the outboard CXRS measurement,  $\Lambda_z^{(n+1)}\equiv \Lambda_i^{(n+1)} + \Lambda^{(n)} = \Lambda_z^{\rm Out}$. In practice only one iteration is necessary because the impurity return flow $\Lambda$ is not very sensitive to the ion parallel flow $\Lambda_i$ and the outboard measurement locations happen to be close to a stagnation point of the calculated impurity return flow, so that $\Lambda_i = \Lambda_z^{\rm Out}$ is already a good guess. 

\begin{figure}[hbt]
\centering
\includegraphics[width=6 cm]{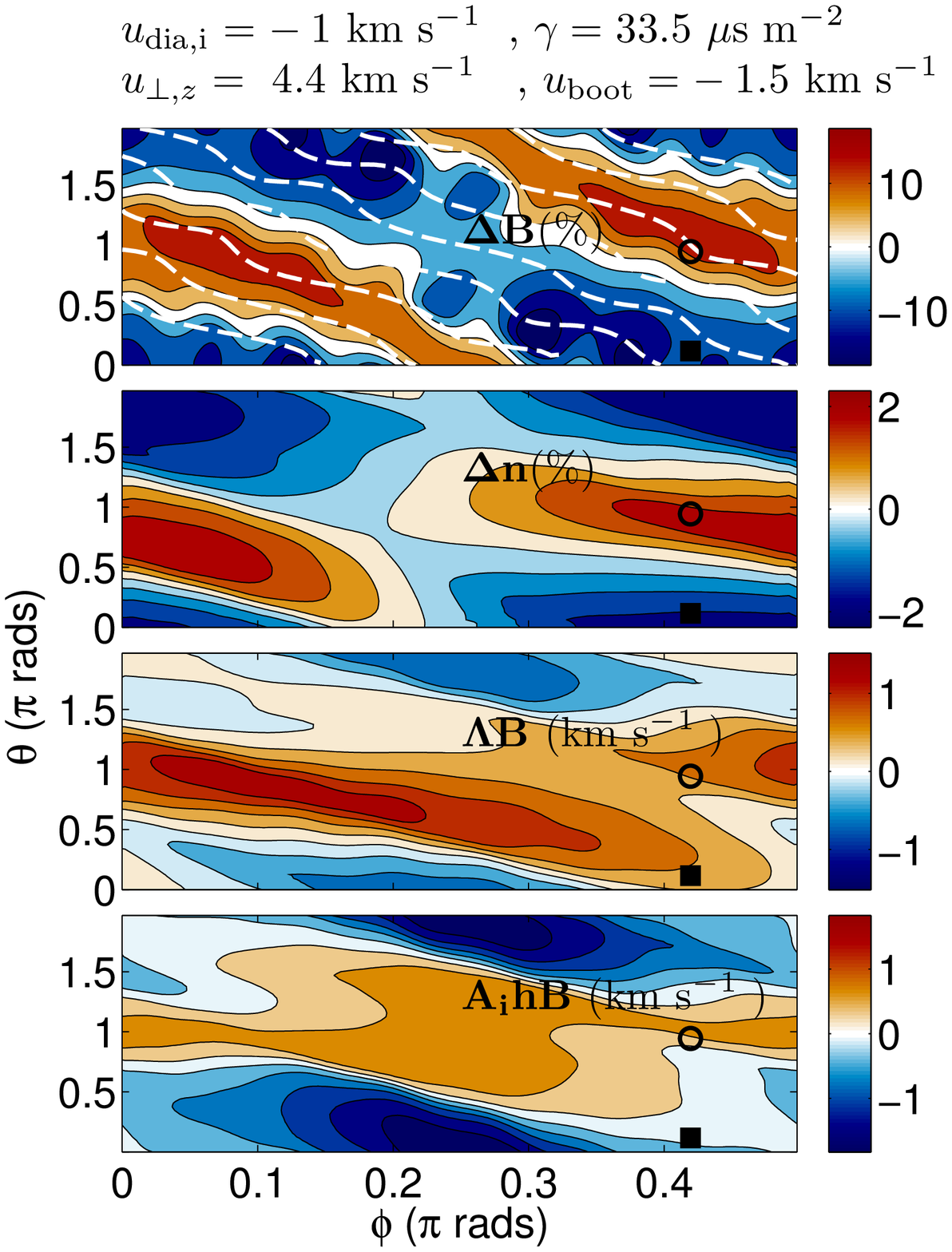}
\includegraphics[width=6 cm]{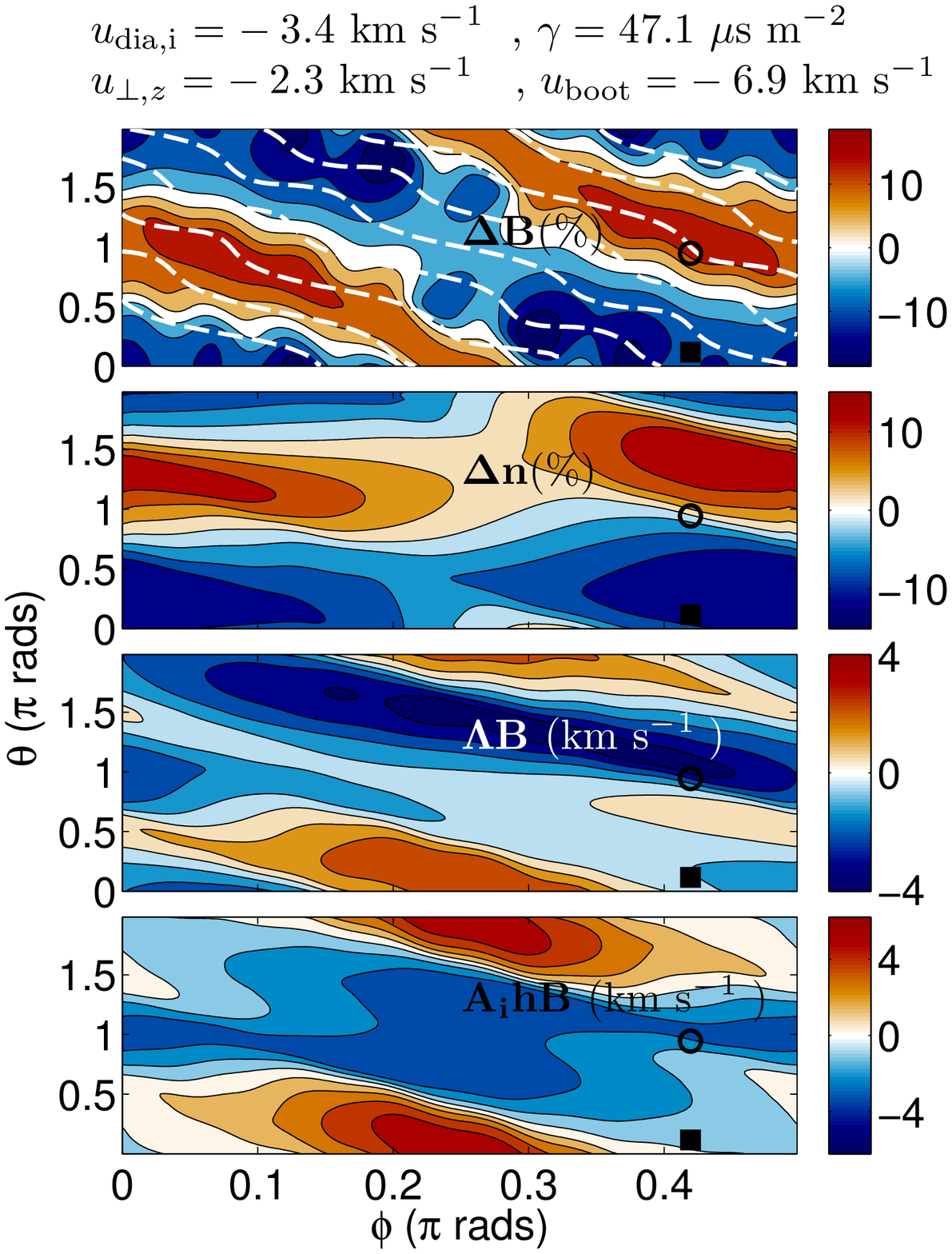}
\caption{\label{fig:density_simul}C$^{6+}$ impurity density redistribution simulation for the discharges $\#25801$ (left) and $\# 32577$ (right) on the surface $\rho=0.6$. From top to bottom: magnetic field strength variation, $\Delta B= B/\langle B\rangle-1$ (field lines are plotted in white); density variation, $\Delta n= n_z/\langle n_z\rangle-1$; return parallel flow, $\Lambda B$; and the differences in the Pfirsch-Schl\"uter velocity, $\Delta u_{\parallel}^{PS}\sim A_ihB$.  The inboard/outboard toroidal measurement positions are shown as an open circle and filled square, respectively.}
\end{figure}
An example of the solution is shown in figure \ref{fig:density_simul} for discharge $\# 32577$ presented in figures \ref{fig:time_evolution} and \ref{fig:plasma_profiles} and for the $\rho=0.6$ magnetic surface. These results correspond to fully ionised carbon C$^{6+}$ impurity that is used for the CXRS measurements. From top to bottom the relative variations of magnetic field strength $B$ and impurity density, the corresponding impurity return flow and the parallel friction drive $A_i hB$ are plotted.  The inboard/outboard toroidal measurement positions are shown as an open circle and solid square, respectively. As a reference the same quantities obtained for the low-density ECRH heated discharge $\#25801$ are presented. The slightly hollow density profiles in typical ECRH discharges in TJ-II results in a small and negative thermodynamic force $A_i$, see equation \eref{eq:friction_centrif_coeff}. Correspondingly, both the relative impurity density variations and return flow are small. Plasma profiles and CXRS flow measurements for this discharge can be found in reference~\cite{ArevaloNF2013}. In particular we recall that the measured impurity flows were shown to be nearly incompressible for this discharge.

\begin{figure}[htb]
\centering
\includegraphics[width=5 cm]{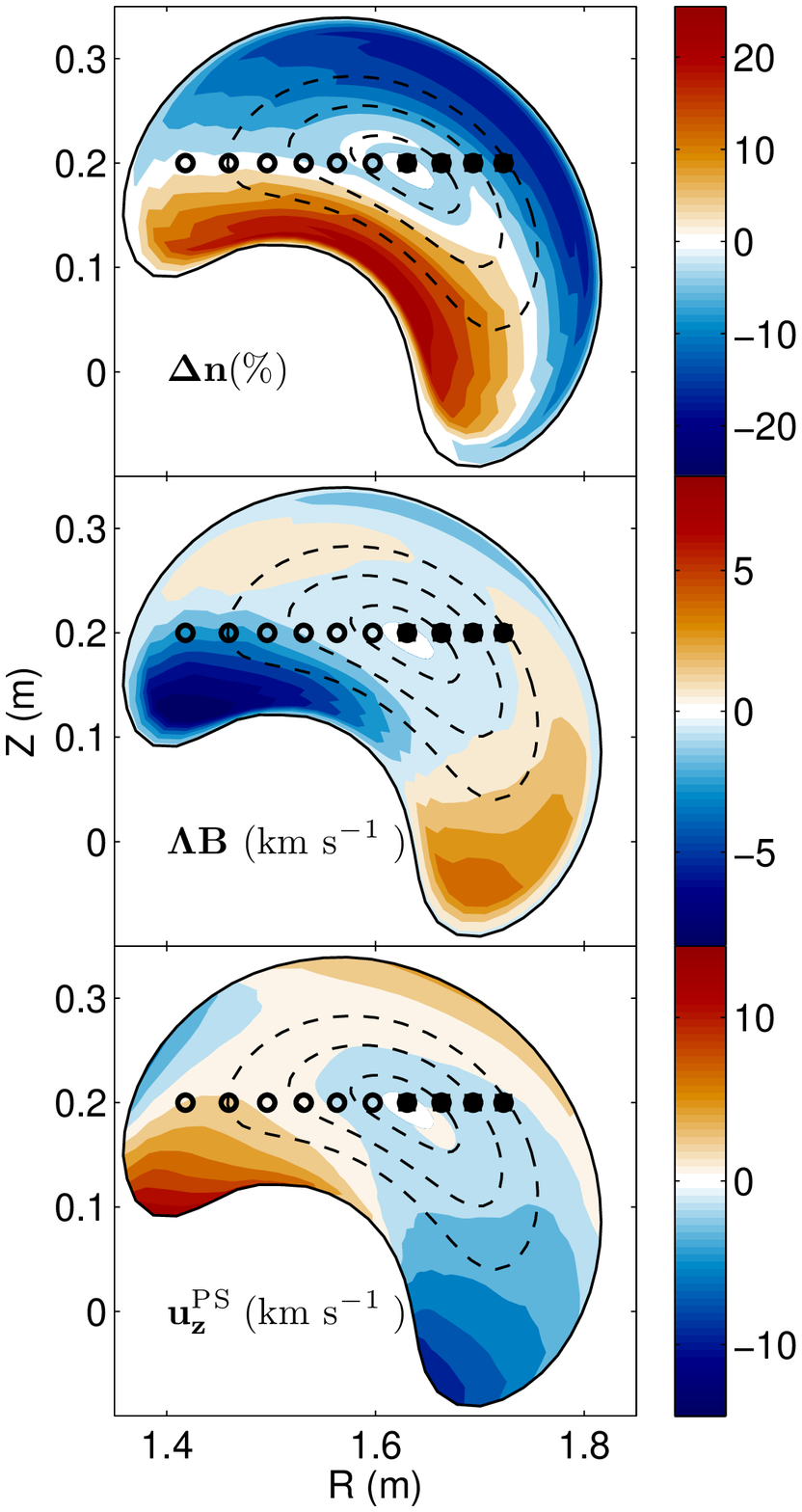}
\caption{\label{fig:toroidal_density_simul}Mapping of C$^{6+}$ impurity density redistribution simulation in the CXRS poloidal plane of measurement for the discharge $\# 32577$. From top to bottom: density variation, $\Delta n= n_z/\langle n_z\rangle-1$; impurity return flow, $\Lambda B$; and the impurity Pfirsch-Schl\"uter velocity. %Regions of stagnation are shown in white, i.e. $n=1$, $\Lambda_zB=0$ and $u^{PS}_{z}=0$. 
The inboard and outboard toroidal measurement positions are shown as open circles and filled squares, respectively.}
\end{figure} 

The results of the calculations of impurity density redistribution for discharge $\# 32577$ and for several magnetic surfaces, $\rho\leq0.8$, are plotted in figure \ref{fig:toroidal_density_simul} for the toroidal section of the CXRS measurements, $\phi=75.5^{\rm o}$. Again, the inboard and outboard toroidal measurement positions are shown as open circles and filled squares, respectively. The magnetic surfaces in which the inboard/outboard comparison is made, namely $\rho\sim0.2,0.4$ and $0.6$, are also shown as dashed lines.  The first two graphs of the figure are the normalized impurity density redistribution, $\Delta n= n_z/\langle n_z\rangle-1$, and impurity return flow, $\Lambda B$. The Pfirsch-Schl\"uter impurity flow, ${u}_z^{\rm PS}=E_zhB$, is presented at the bottom of the figure. 

Some general comments on the solution can be made in light of the simulation results shown in figures \ref{fig:density_simul} and \ref{fig:toroidal_density_simul}. For the plasma profiles of the database used in this work, $\bar{n}_{\rm e}\!\in\!(1.2\!-\!2.4)\times\! 10^{19}$ m$^{-3}$, carbon impurities tend to accumulate in the interior region of the bean-shaped plasma poloidal cross section (which is close the region of maximum magnetic field strength in TJ-II due to the proximity of the central coil). The resulting return flow is comparable in size to the PS impurity flow. Its angular dependence also shows a dominant $\cos{\theta}$ component. The difference in sign between the PS and return parallel flows is in line with the overall tendency heuristically described at the beginning of this section: the return flow $\Lambda$ tends to compensate the $A_ihB$ friction drive in equation~\eref{eq:parallel_momentum_simplified}, for in these ion-root plasmas $A_i$ and ${\rm d}\Phi/{\rm d}\rho$ are of similar magnitude and different signs so that $u_z^{PS}\approx ({\rm d}\Phi/{\rm d}\rho) hB \sim -A_ihB$. Consequently, the differences in the calculated impurity parallel return flow at the locations of the CXRS measurements, $\Delta \Lambda_z\equiv\Delta\Lambda$ from \eref{eq:compressible_flow}, are found to be negative for the ion-root plasma discharges in our CXRS database.
%--------------------------------------------
\subsection{Comparison with experiment}
%--------------------------------------------
\begin{figure}[htb]
\centering
\includegraphics[width=7 cm]{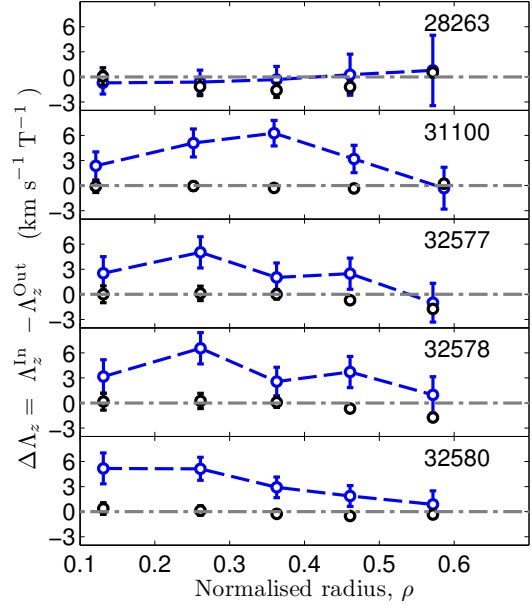}
\caption{\label{fig:friction_failure_0}Radial profiles of the friction-driven simulated (black) and experimental (blue) differences in the parallel mass flow $\Delta\Lambda_z$, for the same discharges presented in figure~\ref{fig:plasma_profiles}.}
\end{figure}

\begin{figure}[htb]
\centering
\includegraphics[width=7 cm]{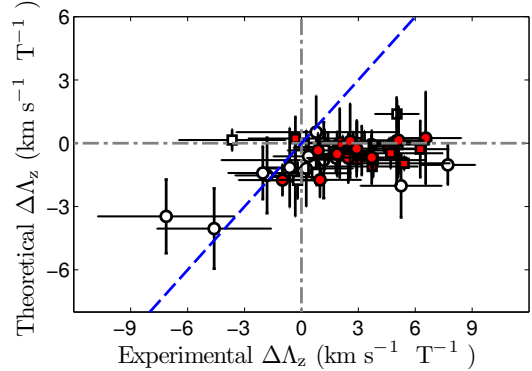}
\caption{\label{fig:friction_failure}Comparison of the experimental and theoretical values of $\Delta \Lambda_z$. It is observed that the expected and measured values are systematically in disagreement, except for a few values. The symbols legend is the same as in figure~\ref{fig:Er_nbi}. The error bars in the simulation come from the spread of the calculated velocities in the measurement volumes.}
\end{figure}

Figure~\ref{fig:friction_failure_0} shows the radial profiles of the friction-driven simulated (black) and experimental (blue) differences in the impurity parallel mass flow, $\Delta\Lambda_z$, for the discharges presented in figure~\ref{fig:plasma_profiles}. The calculated compressible modifications to the impurity flow do not account for the observed differences. A comparison of the experimental and theoretical values of $\Delta \Lambda_z$ is presented in figure~\ref{fig:friction_failure} for the same database as in figure~\ref{fig:Er_nbi}. Error bars in figures \ref{fig:friction_failure_0} and \ref{fig:friction_failure} come from the spread of the calculated velocities in the measurement volumes. 
As discussed in section~\ref{sec:simulation_results}, the parallel friction term in equation \eref{eq:parallel_momentum_simplified} calculated from experimental profiles appears capable of producing a measurable impurity density asymmetry and parallel return flow, even for the internal positions considered in this work (the region of maximum gradient is typically located at $\rho\sim0.7-0.8$ in TJ-II plasmas). Values of $\Delta \Lambda_z^{\rm theo} \sim -2~{\rm km~s^{-1}~T^{-1}}$, or larger, are found in the simulation, while the  experimental differences can easily  reach $\Delta \Lambda_z^{\rm exp} \sim 6~{\rm km~s^{-1}~T^{-1}}$. The overall tendency of the calculated return flows to be more negative in the inboard side is also clear from figure \ref{fig:friction_failure}. Note that at the inboard positions the impurity return flow varies sharply, see figure~\ref{fig:density_simul}, which translates into large error bars. 

From this comparison it is concluded that, whereas the impurity-ion parallel friction (in its model form in equation \eref{eq:parallel_momentum_simplified}) is capable of causing impurity density asymmetries and return flows of the order of magnitude of the observed in-out flow differences, the calculated return flows do not agree with the observed in-surface variation of the impurity parallel mass flow $\Lambda_z$ at the locations of the CXRS measurements for most cases. In the following section some of the assumptions made in the model \eref{eq:parallel_momentum_simplified} are examined, and the parallel force balance \eref{eq:impurity_mom_balance} is extended to account for the impurity inertia and the effect of a parallel electric field.

%%%%%%%%%%%%%%%%%%%%%%%%%%%%%%%%%%%%%%%%%%%%%%%%%%%%%%%%%%%%%%%%%%%%%%%%%%%%%%%%%%%%%%%%%%%%%%%%%%
\section{Discussion on the validity and extensions of the model}\label{sec:extensions}
%%%%%%%%%%%%%%%%%%%%%%%%%%%%%%%%%%%%%%%%%%%%%%%%%%%%%%%%%%%%%%%%%%%%%%%%%%%%%%%%%%%%%%%%%%%%%%%%%%
Previous impurity parallel friction models for stellarators~\cite{Braun_Helander_ConfSeries2010} consider main ions in the Pfirsch-Schl\"uter regime. This regime is not strictly applicable to the plasmas presented here ($n_{\rm i}\in(0.5-3)\times 10^{19}$ m$^{-3}$, $T_{\rm i}\in 100-200$ eV) since main ions are in the plateau regime $\hat{\nu}_{ii}\sim 10^{-1}-10^0$~\cite{VelascoPPCF2012}. As indicated in section \ref{sec:discussion} and explained in the \ref{sec:appendix}, no assumption is made in this work on bulk ion collisionality, although the main ion distribution function is truncated in the Laguerre expansion ($j\leq 1$, see \ref{sec:appendix}) as in the $13$ M approximation \cite{HelanderCollisions,SugamaPoP2010}.
\begin{figure}[htb]
\centering
\includegraphics[width=6 cm]{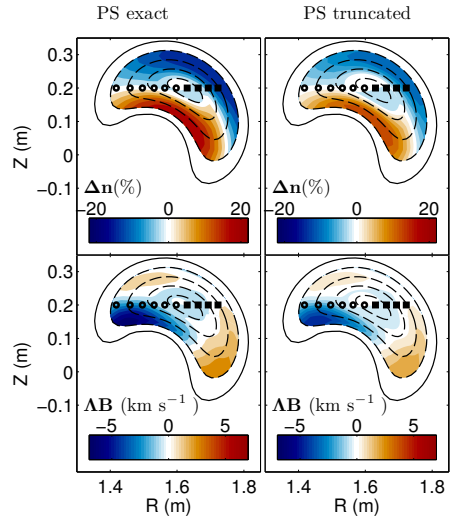}
\caption{\label{fig:toroidal_density_simul_coll}Mapping of C$^{6+}$ impurity density redistribution and return flow  in the CXRS polodial plane of measurement for the discharge $\# 32577$. On the left, the exact Pfirsch-Schl\"uter main ion distribution function is used \cite{Braun_Helander_ConfSeries2010}.%, which results in $B_i^{\rm PS}=0$ and $A_i^{\rm PS} =(T_i/e)\times ({\rm d}\ln n_i /{\rm d}\rho)$.
 On the right, the $j>1$  truncation is applied to the PS exact result.% which yields $A_i^{\rm PS, truncated} =A_i^{\rm PS, truncated}-1/(2e)\times({\rm d}T_i /{\rm d}\rho)$.
}
\end{figure} 

In order to quantify the impact of this approximation let us consider main ions in the  Pfirsch-Schl\"uter regime, as in reference \cite{Braun_Helander_ConfSeries2010}. In this regime of collisionality $\langle{\bf q}_i\cdot{\bf B}\rangle=0$ (hence $B_i=0$ in \eref{eq:thermodynamic_force_B}) and $A_i^{\rm PS} =(T_i/e)\times ({\rm d}\ln n_i /{\rm d}\rho)$. Now, if the $j>1$ truncation is applied to the exact collisional result, the thermodynamic force $A_i$ in \eref{eq:parallel_momentum_simplified} results 
\[
A_i^{\rm PS,~truncated}\!=\!A_i^{\rm PS}-\frac{1}{2e}\frac{{\rm d}T_i}{{\rm d}\rho}, 
\]
which equals the general result $A_i$ in \eref{eq:thermodynamic_force_B}. The resultant impurity redistribution and return flow as obtained from the exact and truncated collisional results are displayed in the left and right columns of figure \ref{fig:toroidal_density_simul_coll}, respectively. As observed, the impurity density in-surface variation reaches values up to $\Delta n\sim\pm 20~\%$ in the exact collisional result while $\Delta n\sim\pm 13~\%$ is found when truncating the main ion distribution function. The simulated return velocity, $\Lambda B$, is similarly affected by the truncation (values of $\pm6$ and $\pm4$ km s$^{-1}$ are obtained in the exact and truncated friction models, respectively). The comparison in figure \ref{fig:toroidal_density_simul_coll}, and the proximity of TJ-II main ion collisionalities to the Pfirsch-Schl\"uter regime, indicate that  the inclusion of higher order Legendre components \cite{SugamaPoP2010} in the modelled friction \eref{eq:parallel_momentum_simplified} is unlikely to change the tendencies in the simulated  impurity redistribution and return flow presented in section \ref{sec:discussion}. On the other hand, the generalization of the parallel friction presented in \eref{eq:parallel_momentum_simplified} allows us to directly use the measured main-ion parameters (since no assumption is made on collisionality) and to include the effect of a non-zero parallel heat flow, thus extending previous friction models in stellarators \cite{Braun_Helander_ConfSeries2010}.

Besides the above discussion on the generalization of the ion-impurity parallel friction, the impurity parallel force balance \eref{eq:impurity_mom_balance} can be extended to account for  inertial and electrostatic parallel forces as
\begin{equation}
\label{eq:impurity_mom_balance_complete}
m_zn_z{\bf b}\cdot({\bf u}_z\cdot\nabla{\bf u}_z)+n_zZe\nabla_\parallel{\Phi} + T_z\nabla_\parallel n_z = R_{z\parallel},
\end{equation}
where the first term is the impurity parallel inertia and the second one is the parallel electric field. The former can be approximated by 
\[
m_zn_z{\bf b}\cdot({\bf u}_z\cdot\nabla{\bf u}_z)\approx
m_zn_z\Lambda_i^2{\bf B}\cdot\nabla B,
\]
since the local PS and return parallel flows are expected to be smaller than the main ion parallel mass flow in internal regions $|\rho|\leq0.4$ of TJ-II NBI heated plasmas. This same approximation leads to the centrifugal outboard accumulation of high-$Z$ impurities in tokamaks \cite{ReinkePPCF2012}. In order to examine the impact of the inertia on the impurity density redistribution, let the impurity parallel force balance be dominated by the inertia, i.e. $\nabla_\| \ln n_z = -\gamma_c^2\nabla_\|b^2$ with $b\equiv B/\langle B\rangle$, $\gamma_c(\rho)\equiv \Lambda_i\langle B\rangle/v_z$ and $v_z=\sqrt{2T_z/m_z}$ the impurity thermal velocity. Then the impurity inhomogeneity is $\Delta n=\exp\left(-\gamma_c^2[b^2-1]\right)$. Note that  although main ion parallel mass flows $\leq 20$ km s$^{-1}$ are comparable to thermal velocities $\leq 50$ km s$^{-1}$  (i.e. $\gamma_c\leq 0.4$), the large aspect ratio of TJ-II $(b^2-1)\sim a/R\sim0.1$ makes $\Delta n\leq 2~\%$ for all the plasma minor radius (here, $a\!\sim\!0.2$ m and $R\!=\!1.5$ m are the minor and major plasma radius). Such an estimation has been confirmed numerically. Therefore, the impurity inertia is neglected henceforth.

\begin{figure}[htb]
\centering
\includegraphics[width=7 cm]{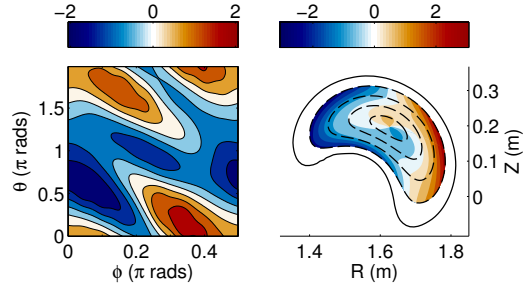}
\caption{\label{fig:toroidal_potential_variation} Simulation of the electrostatic potential inhomogeneity, $\tilde\Phi$ in Volts, for discharge $\#32577$. (Left) In-surface variations for $\rho=0.6$. (Right) Mapping in the CXRS poloidal plane for several magnetic surfaces, $\rho\leq0.8$. 
}
\end{figure}

On the other hand,  the term containing the electrostatic potential variation in the flux surface, $\tilde{\Phi}=\Phi-\langle\Phi\rangle$, in equation \eref{eq:impurity_mom_balance_complete} is considered. This portion of the full electrostatic potential, $\Phi$, results from imposing quasi-neutrality among the non-equilibrium density pieces of the coexistent species. Furthermore the calculation, carried out with the particle in cell code EUTERPE~\cite{Regana2012}, considers adiabatic electron response and trace impurities. Under these approximations the resulting map of $\tilde{\Phi}$ mirrors that of the main ion density. As an example, $\tilde{\Phi}$ is shown in figure \ref{fig:toroidal_potential_variation} for the discharge $\# 32577$. On the left column, $\tilde{\Phi}$ at the surface $\rho=0.6$ is represented, while on the right $\tilde\Phi$ is displayed at several magnetic surfaces, $\rho\leq0.8$, for the toroidal section of the CXRS measurements.

\begin{figure}[htb]
\centering
\includegraphics[width=7 cm]{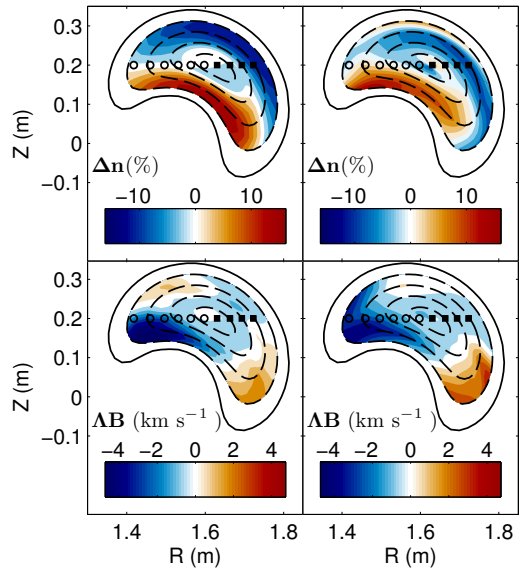}
\caption{\label{fig:force_comparison} 
Mapping in the CXRS poloidal plane of the simulated C$^{6+}$ density inhomogeneity, $\Delta n$, and return flow, $\Lambda {\bf B}$, for discharge $\#32577$ and for several magnetic surfaces, $\rho\leq0.8$, after considering: (left) friction only and (right) all the relevant  forces in model \eref{eq:parallel_momentum_simplified_complete}.
}
\end{figure}

Thus, if parallel inertia is neglected, the impurity momentum balance \eref{eq:impurity_mom_balance_complete} results as
\begin{equation}\label{eq:parallel_momentum_simplified_complete}
{\bf B}\cdot\nabla\ln n=\gamma_f B^2\left(A_ih+B_i-\Lambda\right)-\frac{eZ}{T_z}{\bf B}\cdot\nabla\Phi,
\end{equation}
The parallel momentum balance in its form \eref{eq:parallel_momentum_simplified_complete}, together with particle conservation \eref{eq:continuity_equation_simplified}, can be transformed into a second order partial differential equation for the unknown $n$, as in section \eref{sec:simulation_results}. Figure \ref{fig:force_comparison} displays a mapping in the CXRS poloidal plane of measurement of the simulation results for discharge $\#32577$ and for several magnetic surfaces, $\rho\leq0.8$, after considering (left) only friction and (right) friction plus the $\nabla_\|\Phi$ forces in its model form \eref{eq:parallel_momentum_simplified_complete}. As observed, the impurity redistribution and return flow patterns are affected by the inhomogeneity of the potential only at external radial locations $\rho>0.7$. Nevertheless the tendency to display $\Delta\Lambda<0$ remains unaltered, thus contradicting the experimental observations. 

Finally, a possibly important omission of the impurity re-distribution model used in this work could be the assumption of trace impurities. For the plasmas considered here values of $Z_{\rm eff}\sim 1.2-1.6$ are obtained from soft X-ray emission,  which would give rise to impurity strengths of $n_zZ^2/n_i\sim0.2-0.6$. In such a case both the inhomogeneity of the electrostatic potential \cite{Landreman_PoP2011} and the collision operator used to model the parallel friction on the impurities \cite{ViezzerEPS2013} would change, thus modifying the impurity redistribution within a magnetic surface. The inclusion of these effects is out of the scope of this paper and is left to future work.

%%%%%%%%%%%%%%%%%%%%%%%%%%%%%%%%%%%%%%%%%%%%%%%%%%%%%%%%%%%%%%%%%%%%%%%%%%%%%%%%%%%%%%%%%%%%%%%%%%
\section{Conclusions}
%%%%%%%%%%%%%%%%%%%%%%%%%%%%%%%%%%%%%%%%%%%%%%%%%%%%%%%%%%%%%%%%%%%%%%%%%%%%%%%%%%%%%%%%%%%%%%%%%%
\label{sec:conclusions}
In this work fully-ionised carbon impurity flows in ion-root, NBI heated, TJ-II plasmas are studied by means of Charge Exchange Recombination Spectroscopy. Perpendicular flows are found to be in reasonable agreement with neoclassical calculations of the radial electric field. The parallel flow of the impurity is obtained at two locations on the same flux surface and the calculated Pfirsch-Schl\"uter parallel velocity is subtracted. The remaining component of the flow is systematically observed to vary on each flux surface, pointing to a breakdown of impurity flow incompressibility in the medium density plasmas studied. The experimentally observed velocity deviations are compared with the parallel return flow calculated from a modelled impurity density redistribution driven by ion-impurity friction. Such a model is extended to account for impurity inertia and inhomogeneities in the electrostatic potential. The simulation results show that the parallel impurity force balance is dominated by parallel friction for the plasmas considered here, and demonstrate that the calculated return flow substantially modifies the incompressible velocity pattern. However, these modifications become small at the precise locations of the CXRS measurements and do not explain the in-surface variations of impurity parallel flow. The experimental validation of theoretical models of impurity density redistribution within a flux surface is of considerable importance as it provides an indirect validation of the model predictions for impurity radial transport.  

%%%%%%%%%%%%%%%%%%%%%%%%%%%%%%%%%%%%%%%%%%%%%%%%%%%%%%%%%%%%%%%%%%%%%%%%%%%%%%%%%%%%%%%%%%%%%%%%%%
\section*{Acknowledgement}
%%%%%%%%%%%%%%%%%%%%%%%%%%%%%%%%%%%%%%%%%%%%%%%%%%%%%%%%%%%%%%%%%%%%%%%%%%%%%%%%%%%%%%%%%%%%%%%%%%
The authors are indebted to the TJ-II experimental group.
They would like to thank M. A. Ochando and C. Hidalgo for his support. 
J. Ar\'evalo acknowledges financial support from the FPI grant awarded by CIEMAT 
(BOE resolution nº 171, 24/06/2008). K. J. MCarthy and J. Ar\'evalo acknowledge financial support from the Spanish Ministry of Science and Innovation (ENE2010-19676). M. Landreman was supported by the Fusion Energy Postdoctoral Research Program administered by the Oak Ridge Institute for Science and Education.

%%%%%%%%%%%%%%%%%%%%%%%%%%%%%%%%%%%%%%%%%%%%%%%%%%%%%%%%%%%%%%%%%%%%%%%%%%%%%%%%%%%%%%%%%%%%%%%%%%
\appendix
\section{Impurity density redistribution in a stellarator}\label{sec:appendix}
\setcounter{section}{1}
%%%%%%%%%%%%%%%%%%%%%%%%%%%%%%%%%%%%%%%%%%%%%%%%%%%%%%%%%%%%%%%%%%%%%%%%%%%%%%%%%%%%%%%%%%%%%%%%%%
In this appendix the parallel momentum balance \eref{eq:parallel_momentum_simplified_complete} is derived and the method used to solved this equation consistently with particle number conservation is detailed. The impurity parallel momentum equation is taken to be
\begin{equation}\label{eq:parallel_momentum_equation_appendix}
T_z\nabla_\parallel n_z+n_zZe\nabla_\parallel \Phi=R_{z\parallel},
\end{equation}
where $R_{z\parallel}$ is the parallel friction on the impurities  and $\Phi$ the electrostatic potential. 
As demonstrated in reference \cite{Helander_PoP1998} the impurity temperature is equilibrated with the bulk ion temperature and is therefore constant on the flux surface. As it is also shown in \cite{Helander_PoP1998}, impurity parallel inertia and viscosity can be neglected in \eref{eq:parallel_momentum_equation_appendix} if $\delta_i/(Z\hat{\nu}_{ii})\ll 1$. For the plasmas considered in this work ($n_{\rm i}\in(0.5-3)\times 10^{19}$ m$^{-3}$, $T_{\rm i}\in 100-200$ eV) typical values of the normalised ion gyro-radius are $\delta_i\sim (5-10)\times10^{-3}$. Bulk ions in NBI-heated TJ-II plasmas are in the plateau regime,  $\hat{\nu}_{ii}\!\sim\!10^{-1}\!-\!10^0$, although close to the Pfirsch-Schl\"uter regime of collisionality \cite{VelascoPPCF2012}. Then, for fully-ionised carbon impurity ions ($Z=6$), $\delta_i/(Z\hat{\nu}_{ii})\!\sim\!10^{-3}\!-\!10^{-2}$. Hence, the assumptions made in reference \cite{Helander_PoP1998} to derive equation \eref{eq:parallel_momentum_equation_appendix} are applicable in this work. Furthermore, the expected variations of C$^{6+}$ density within the surface are $\tilde{n}_z/\langle n_z\rangle\sim\delta_i\hat{\nu}_{ii}Z^2\sim 0.1-0.5$, thus justifying the present study.

In the trace impurity limit, $\sum n_zZ^2\ll n_i$, the parallel friction on the impurities may be approximated by~\cite{Landreman_PoP2011}
\begin{equation}
R_{z\parallel}\approx R_{zi\parallel}= -R_{iz\parallel} = -\int {\rm d}^3vm_i v_\parallel C_{iz}\big{\{}f_{i1}\big{\}},
\end{equation}
with $f_{i1}$ the first order departure of the bulk ion distribution function from a Maxwellian. The ion-impurity collision operator consists of a Lorentz operator plus a term guaranteeing momentum conservation \cite{HelanderCollisions}
\begin{eqnarray}
&C_{iz}\big{\{}f_{i1}\big{\}}=\nu_{iz}\mathcal{L}\big{\{}f_{i1}\big{\}}+\nu_{iz}\frac{m_iv_\parallel u_{z\parallel}}{T_i}f_{i0},&\\
&\mathcal{L}=\frac{1}{2}\frac{\partial}{\partial_\xi}\left[(1-\xi^2)\frac{\partial}{\partial_\xi}\right].&
\end{eqnarray}
Here, $\nu_{iz}=3\pi^{1/2}/(4\tau_{iz}x_i^3)$, %$\tau_{iz}=3(2\pi)^{3/2}\varepsilon_0^2m_i^{1/2}T_i^{3/2}/(n_ZZ^2e^4\ln{\Lambda})$
$\tau_{iz}=\tau_{ii}n_i/(n_zZ^2)$ is the ion-impurity collision time, $x_i=v/v_i$, $v_i=\sqrt{2T_i/m_i}$ the ion thermal speed, $\xi=v_\parallel/v$ the pitch-angle   and $f_{i0}=n_{i0}/(\pi^{3/2}v_i^3)\exp\left(-x_i^2\right)$ is a flux-function Maxwellian. Since the collision operator is self-adjoint and $\mathcal{L}\{v_\parallel\}=-v_\parallel$, the term in the parallel friction force arising from the Lorentz operator is written as
\begin{equation}\label{eq:landreman_generalization}
-\int {\rm d}^3v m_iv_\parallel\nu_{iz} \mathcal{L}\big{\{}f_{i1}\big{\}}=
\frac{3\pi^{1/2}}{4\tau_{iz}}m_iv_i\int {\rm d}^3v \frac{\xi}{x_i^2}f_{i1}.
\end{equation}

Let us consider now the expansion of $f_{i1}({\bf x},v,\xi)$ in Legendre polynomials $P_l(\xi)$ $\left[ P_0=1,~P_1=\xi,~{\rm etc.}\right]$
\cite{SugamaPoP2002}. Thanks to the orthogonality properties of the $P_l$ polynomials only the $l\!=\!1$ component of $f_{i1}$ contributes to equation \eref{eq:landreman_generalization}. Such component is associated with the parallel particle and heat flows ($u_{i\|}$ and  $q_{i\|}$, respectively) and is expanded by Laguerre (Sonine) polynomials $L_j^{(3/2)}(x_i^2)$ $\left[ L_0^{(3/2)}=1,~L_1^{(3/2)}=-x_i^2+5/2,~L_1^{(3/2)}=x_i^4/2-7x_i^2/2+15/8,~{\rm etc.}\right]$ as \cite{SugamaPoP2002,HelanderCollisions}
\begin{equation}\label{eq:f_i1_expansion}
f_{i1}^{(l=1)}=\frac{2}{v_{i}}\xi x_i\left\{u_{i\parallel}-L_1^{(3/2)}(x_i^2)\frac{2}{5}\frac{q_{i\parallel}}{p_i}\right\}f_{i0}+f_{i1}^{(l=1,j\ge 2)}.
\end{equation}
Here, $f_{i1}^{(l=1,j\ge 2)}$ denotes the sum of the $j$th Laguerre polynomial components with $j\geq2$. The inclusion of $j>1$ terms \cite{SugamaPoP2010} is out the scope of this paper and thus $f_{i1}^{(l=1)}\approx f_{i1}^{(l=1,j\leq1)}$ is taken in equation \eref{eq:f_i1_expansion}, as it is customary in the moments approach to neoclassical transport \cite{HelanderCollisions} (the so-called $13$ M approximation). See the comments in section \ref{sec:extensions} regarding the effect of this truncation. With this assumption the parallel friction on the impurities reads
\begin{equation}\label{eq:general_friction}
R_{z\parallel}\approx\frac{m_in_{i0}}{\tau_{iz}}\left(u_{i\parallel}-\frac{3}{5}\frac{q_{i\parallel}}{p_i}-u_{z\parallel}\right),
\end{equation}
with $u_{z\|}$ the impurity ion parallel flow. 
\footnote{Note that if the exact result for main ion distribution function in the Pfirsch-Schl\"uter regime is used \cite{Braun_Helander_ConfSeries2010} 
\[
f_{i1}^{(l=1)}=\frac{2}{v_{i}}\xi x_i\left\{u_{i\parallel}-\left(L_1^{(3/2)}(x_i^2)-\frac{4}{15}L_2^{(3/2)}(x_i^2)\right)\frac{2}{5}\frac{q_{i\parallel}}{p_i}\right\}f_{i0},
\]
the pre-factor $-3/5$ accompanying the parallel heat flow in \eref{eq:general_friction} must be replaced by $-2/5$.}
For simplicity energy exchange is neglected, thus making heat flows incompressible for each particle species, $\nabla\cdot{\bf q}_\alpha\approx 0$. Then the bulk ion parallel heat flow is 
\begin{equation}\label{eq:parallel_heat_flow}
q_{i\parallel}=\frac{5p_i}{2e}\frac{\partial T_i}{\partial\rho}hB+\langle {\bf q}_{i\parallel}\cdot{\bf B}\rangle\frac{B}{\langle B^2\rangle},
\end{equation} 
with the function $h$ defined in section \ref{sec:flow_structure}. Using the general expression for a compressible impurity flow, equation \eref{eq:compressible_flow}, and equation \eref{eq:parallel_heat_flow}, the parallel friction on the impurities is finally recast as
\begin{eqnarray}
R_{z\parallel}&=&\frac{m_in_{i0}}{\tau_{iz}}B\left(\left[E_i-E_z-\frac{3p_i}{2e}\frac{\partial T_i}{\partial\rho}\right]h-\Lambda -\frac{3}{5}\frac{\langle {\bf q}_{i\parallel}\cdot{\bf B}\rangle}{p_i\langle B^2\rangle}\right)\nonumber\\
&\approx&p_z\gamma_fB\left(A_ih+B_i-\Lambda\right),
\end{eqnarray}
with the flux constants $\gamma_f(\rho)$, $A_i(\rho)$ and $B_i(\rho)$ given by equations \eref{eq:flux_const_friction}.
In the last step, the impurity diamagnetic term has been neglected against the main ion one. With these assumptions (i.e. %$\delta_i/(Z\hat{\nu}_{ii})\ll 1$, 
trace impurities, $\sum n_zZ^2\ll n_i$, and $13$ M approximation, $f_{i1}^{(l=1)}\approx f_{i1}^{(l=1,j\leq1)}$) % and $v_{\rm diam,i}\gg v_{\rm diam,z}$)
 the impurity parallel momentum balance \eref{eq:parallel_momentum_equation_appendix} results
\begin{equation}\label{eq:dens_variation}
{\bf B}\cdot\nabla n_z=\gamma_{\rm f} n_zB^2\left(A_ih+B_i-\Lambda \right)-n_z\frac{eZ}{T_z}{\bf B}\cdot\nabla \Phi,
\end{equation}
hence recovering equation \eref{eq:impurity_mom_balance_complete} and the simplified form \eref{eq:parallel_momentum_simplified}, when the inhomogeneity of the potential, $\tilde\Phi\equiv \Phi-\langle\Phi\rangle$, is neglected. In stellarator geometry, equation~\eref{eq:dens_variation} and the continuity equation \eref{eq:continuity_equation} form a coupled system \cite{Braun_Helander_ConfSeries2010} of partial differential equations (PDEs). This set of equations can be expressed as a parabolic PDE in the variable $n=n_z/\langle n_z\rangle$
\begin{equation}
\label{eq:second_order_PDE_n}
{\bf B}\cdot\nabla\left({\bf B}\cdot\nabla n\right)-g{\bf B}\cdot\nabla n-\gamma_{\rm f} B^2{\bf u}_{z\perp}\cdot\nabla n-fn=0,
\end{equation}
where
\begin{eqnarray}
g(\rho,\theta,\phi)&=&{\bf B}\cdot\nabla \ln B^2-\frac{eZ}{T_z}{\bf B}\cdot\nabla \Phi+\nonumber\\
&&\gamma_{\bf f}B^2\left\{(A_i+E_z)h+B_i+\Lambda_i\right\},\\
f(\rho,\theta,\phi)&=&
\gamma_{\rm f} A_i{\bf B}\times\nabla\rho\cdot\nabla\ln{B^2}+\nonumber\\
&&\frac{eZ}{T_z}\left({\bf B}\cdot\nabla \ln B^2-{\bf B}\cdot\nabla\right){\bf B}\cdot\nabla\Phi.
\end{eqnarray}
Equation~\eref{eq:second_order_PDE_n} is converted to an algebraic system of equations by applying finite differences to the variable $n$. The angular periodicity of the TJ-II ($T_\theta\!=\!2\pi$ and $T_\phi\!=\!\pi/2$) and the condition $\langle n\rangle\!=\!1$ are imposed. The  parallel return flow $\Lambda B$ is obtained from equation~\eref{eq:continuity_equation_simplified}. The system of PDEs has also been solved by Fourier expanding the variables in Boozer coordinates, showing consistency with the finite differences scheme.
%%%%%%%%%%%%%%%%%%%%%%%%%%%%%%%%%%%%%%%%%%%%%%%%%%%%%%%%%%%%%%%%%%%%%%%%%%%%%%%%%%%%%%%%%%%%%%%%%%
\section*{References}
%%%%%%%%%%%%%%%%%%%%%%%%%%%%%%%%%%%%%%%%%%%%%%%%%%%%%%%%%%%%%%%%%%%%%%%%%%%%%%%%%%%%%%%%%%%%%%%%%%

\end{document}